\documentclass{LT23auth}
\usepackage{graphicx}

\begin{document}

\begin{frontmatter}

\title{Transport via a Quantum Shuttle}

\author[address1,address2]{Angus MacKinnon\thanksref{thank1}},
\author[address3]{Andrew D. Armour}

\address[address1]{The Blackett Laboratory, Imperial College, London SW7 2BW, UK}

\address[address2]{The Cavendish Laboratory, Madingley Rd, Cambridge CB3 0HE,
UK}

\address[address3]{School of Physics and Astronomy, University of Nottingham,
Nottingham NG7 2RD, UK}

\thanks[thank1]{Corresponding author. Present address:  Condensed Matter
Theory Group, The Blackett Laboratory, Imperial College, London SW7 2BW, UK
 E-mail: a.mackinnon@ic.ac.uk}

\begin{abstract}
We investigate the effect of quantisation of vibrational modes on 
a system in which the transport path is through a quantum dot mounted 
on a cantilever or spring such that tunnelling to and from the dot is modulated
by the oscillation.  We consider here the implications of quantum
aspects of the motion. Peaks in the current
voltage characteristic are observed which correspond to avoided level crossings in
the eigenvalue spectrum.  Transport occurs through processes in which phonons are
created.  This provides a path for dissipation
of energy as well as a mechanism for driving the oscillator, thus making it easier
for electrons to tunnel onto and off the dot and be ferried across the
device.

\end{abstract}

\end{frontmatter}

\begin{figure}[ht]
\begin{center}\leavevmode
\includegraphics[width=\linewidth,]{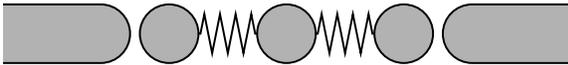}
\caption{A quantum shuttle consisting of a dot on springs flanked
by 2 stationary dots attached to semi--infinite leads.
}\label{fig:shuttle}\end{center}\end{figure}

Recent advances in the fabrication of nanomechanical devices\cite{Cleland02}
are showing a distinct trend towards systems in which a quantum description
is required not only for the electronic behaviour but also for the mechanical aspects. We
therefore consider a model nano--electro--mechanical system consisting
of a quantum dot attached to a spring or cantilever which moves between
2 contacts; thus acting as an electron shuttle (Fig. \ref{fig:shuttle}).
Devices of this sort have previously been fabricated albeit with classical
mechanical behaviour\cite{Gk}.

In this work we consider 2 models:
\begin{enumerate}
\item\label{model1} a 3 dot model in which the moveable dot is flanked by 2 stationary
dots.  This is described by a tight--binding model\cite{AmK02a}.
\begin{eqnarray}
{{H}}&=&\varepsilon_l|l\rangle\langle l|+\varepsilon_r|r\rangle\langle r|
+\varepsilon_c(\hat{x})|c\rangle\langle c|+
\hbar\omega\hat{d}^{\dagger}\hat{d} \\ \nonumber 
&&-V{\mathrm{e}}^{-\alpha(\hat{d}^{\dagger}+\hat{d})}
(|c\rangle\langle l|+| l\rangle\langle c|)\\ \nonumber 
&&-V{\mathrm{e}}^{\alpha [(\hat{d}^{\dagger}+\hat{d})-2x_0]}
(|c\rangle\langle r|+| r\rangle\langle c|),
\label{eq:3dots}
\end{eqnarray}
where $| i\rangle\langle i |$ are projection
operators for the three  electronic states and the vibrational mode, frequency
$\omega$, is operated on by  $\hat{d}$. 

\item\label{model2} a scattering model in which the dot is embedded between semi--infinite
leads in a Landauer\cite{Lan70} geometry.
\begin{eqnarray}
H&=&-E_{\rm e}{\partial^2\over\partial x_{\rm e}^2}
- E_{\rm s}
\left({\partial^2\over\partial y_{\rm s}} 
- {\textstyle\frac14}y^2_{\rm s}\right) \\
&&+ V_1\left[\Xi(x_{\rm e}, d_{\rm c}) - \Xi(x_e-s y_{\rm s},
d_s)\right]\nonumber
\label{eq:scatter}
\end{eqnarray}
where $x_{\rm e}$ and $y_{\rm s}$ represent the electron
and phonon coordinates respectively, $E_{\rm e}$ and $E_{\rm s}$
the corresponding energy scales, $s$ the shuttle displacement,
\[
\Xi(x, d)
= \Theta(x_{\rm e} + {\textstyle\frac12}d) 
- \Theta(x_{\rm e} - {\textstyle\frac12}d)
\]
a barrier of width $d$; $d_{\rm c}$ and $d_{\rm s}$ represent the separation
of the contacts and the size of the shuttle respectively.
\end{enumerate}
The presence of the exponential terms in (\ref{eq:3dots}) and a similar
term in the matching conditions of (\ref{eq:scatter}) make
the tunnelling rates sensitive to the position of the shuttle.

\begin{figure}[ht]
\begin{center}\leavevmode
\hbox{
\includegraphics[width=0.5\linewidth]{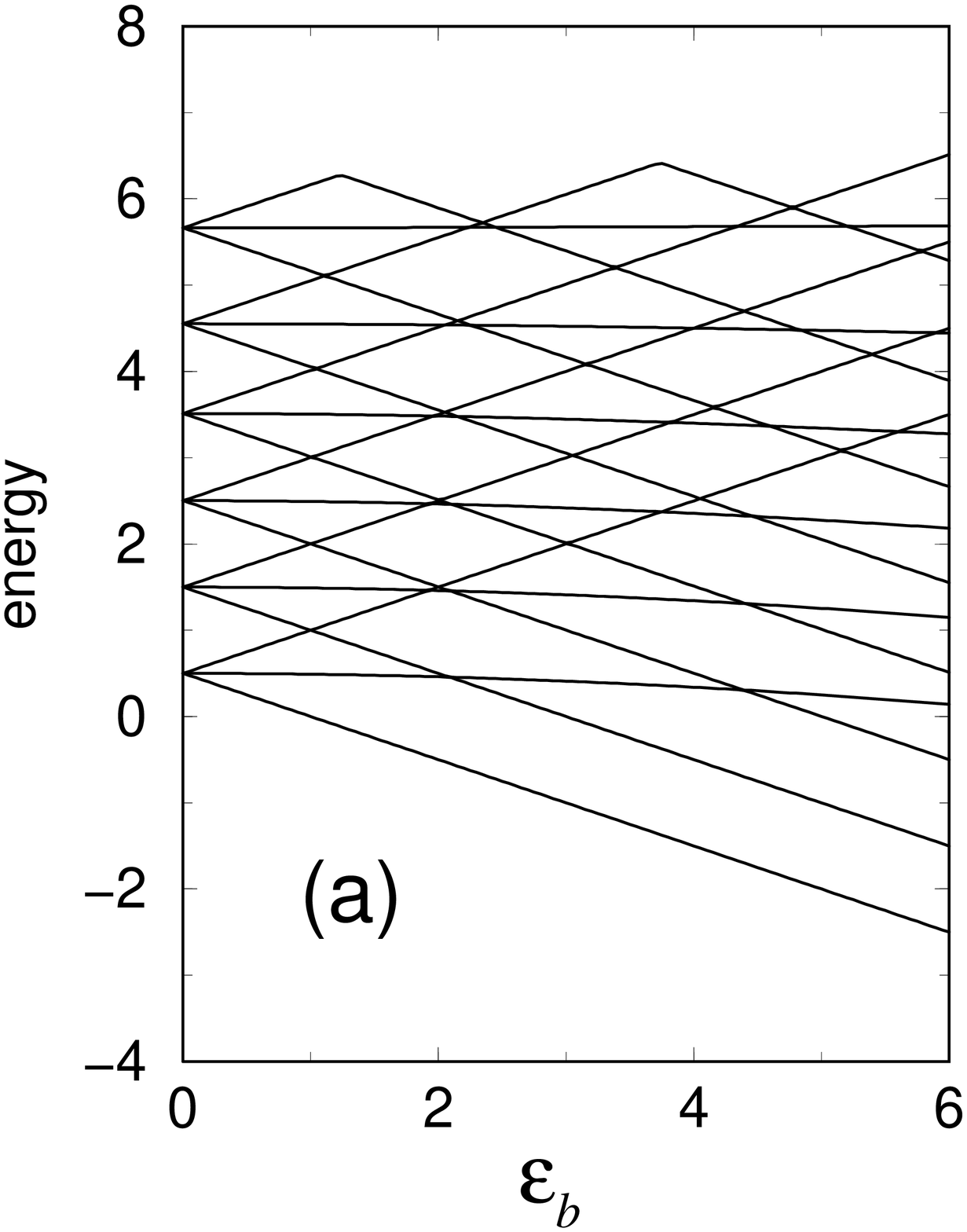}
\includegraphics[width=0.5\linewidth]{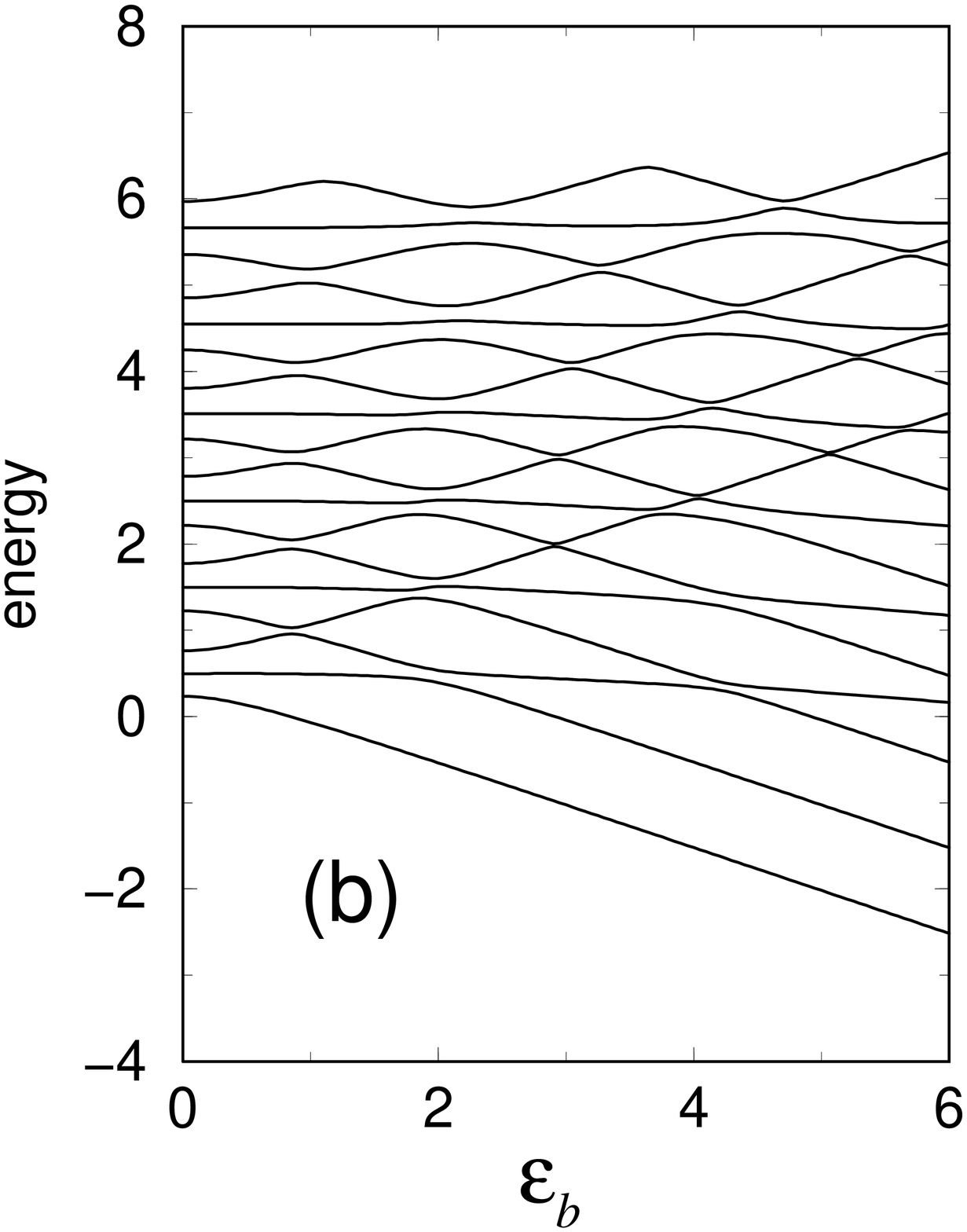}}
\caption{Eigenvalues of the 3--dot model as a function of potential difference
between the outer dots; (a) without tunnelling, (b) with tunnelling.
}\label{fig:anticross}\end{center}
\end{figure}
When the energies of the 3 dot system are plotted against the potential difference
between the outer dots the tunnelling through the shuttle results in a
series of anti-crossings (Fig.~\ref{fig:anticross}).  These are of 2 types:
those involving only the outer dots, such as at $(1,1)$ in fig.~\ref{fig:anticross};
those involving all 3 dots, such as at $(2,1.5)$ in fig.~\ref{fig:anticross}.

\begin{figure}[ht]
\begin{center}\leavevmode
\includegraphics[height=0.4\linewidth]{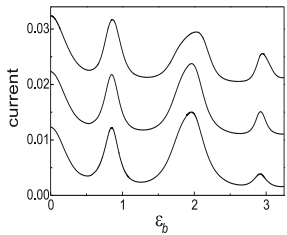}
\caption{Current in the 3--dot model as a function of potential difference
for 3 different damping rates (shifted for clarity).
}\label{fig:sh_gam}\end{center}
\end{figure}
The resulting current--voltage characteristic is shown in fig.~\ref{fig:sh_gam}.
There is a peak at $\epsilon_b=0$ due to resonant tunnelling through all
3 dots.  The other features may be associated with anti--crossings in fig.~\ref{fig:anticross}.
The maximum at $\varepsilon_b\approx 0.8$ occurs when the difference
between the energies of the left and right--hand dots differ by that
of a single phonon, whereas the peak at $\varepsilon_b\approx2.0$ involves
2 phonons and is associated with a 3--way anti--crossing.  
Note, in particular, that the latter peak is much more sensitive
to the damping of the phonons than is the former.  This is clearly due
to the involvement of the state on the shuttle itself.

\begin{figure}[ht]
\begin{center}\leavevmode
\includegraphics[width=\linewidth]{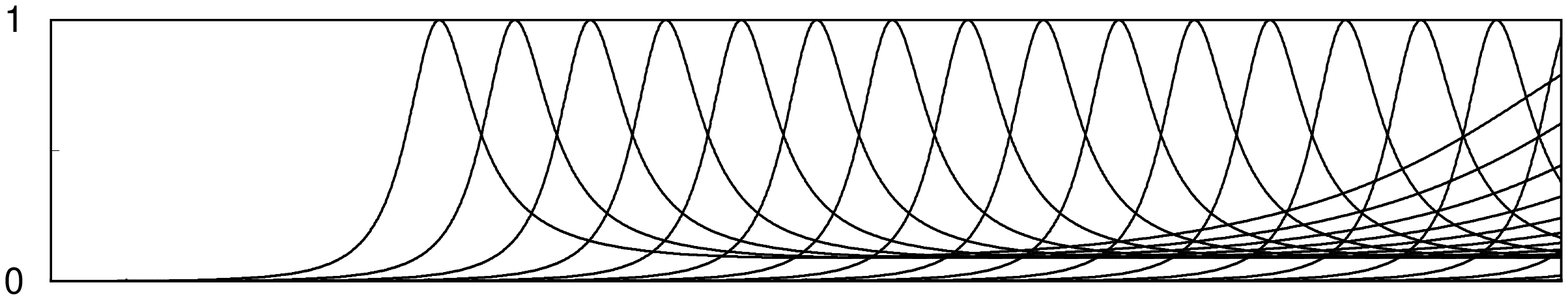}
\includegraphics[width=\linewidth]{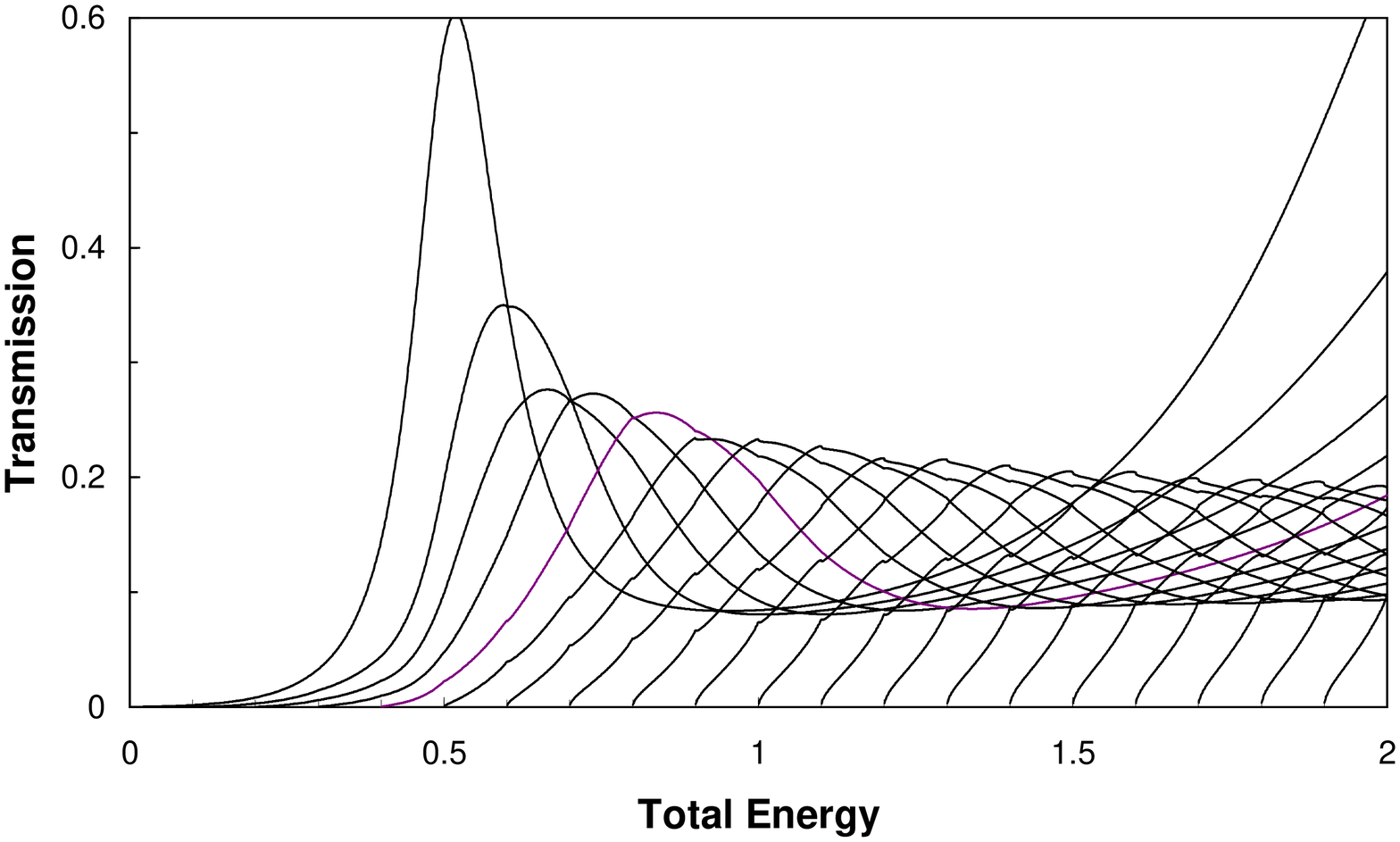}
\caption{Total transmission probability as a function of total energy ($\mbox{electron}+\mbox{phonon}$)
for various incident phonon numbers ($n=0-19$ start at E=n/10). Shuttle displacement is
zero (upper figure), about 40\% of the barrier width (lower figure).  
The onset of a 2nd resonant tunnelling peak is seen to the right.}
\label{fig:scatt_trans}\end{center}
\end{figure}
Results for the scattering model are illustrated in fig.~\ref{fig:scatt_trans}.
The various curves correspond to different states of the shuttle before
the electron is scattered.  Note that the half--width of the peaks settles down
to about double the phonon energy, corresponding to a transmission time
of about half the shuttle period as would be expected for the shuttle effect.

We have presented results for 2 different models of a  quantum shuttle.
In both cases there was no necessity to include a specific mechanism to
drive the shuttle.  As long as the potential difference across the system
is greater than the phonon energy the shuttle may be driven by the creation
of phonons.  This in turn implies that the dissipation of this energy will
play an important role in the behaviour of such a system.

\begin{ack}
We would like to thank the EPSRC for financial support and the Cavendish Laboratory,
Cambridge for its hospitality.
\end{ack}

%
%

\end{document}